\documentclass[]{interact}

\usepackage[normalem]{ulem}
\usepackage{xcolor}
\usepackage{comment}
\usepackage{lmodern}
\usepackage[authoryear]{natbib}
\usepackage{epstopdf}
\usepackage[caption=false]{subfig}

\theoremstyle{plain}
\newtheorem{theorem}{Theorem}

\theoremstyle{definition}

\theoremstyle{remark}
\newtheorem{remark}{Remark}

\begin{document}

\title{Local Quasi-Linear Models: Kernel Differential Equation Regression and Fire Data Analysis}

\author{
\name{Chunlei Ge\textsuperscript{a} and W. John Braun$^\ast$\textsuperscript{a}\thanks{$^\ast$CONTACT W. John Braun. Email: john.braun@ubc.ca}}
\affil{\textsuperscript{a}The University of British Columbia Okanagan, 3333 University Way, Kelowna, BC, Canada}
}

\maketitle

\begin{abstract}
We introduce the local quasi-linear (LQL) model, a differential equation-constrained local polynomial regression framework for the general first-order linear ordinary differential equation (ODE) $g'(x)=a(x)g(x)+b(x)$, extending prior work on differential equation-constrained local polynomial regression (DE-constrained LPR) for the exponential growth model. We derive closed-form DE-constrained local polynomial (DE1-$k$) estimators for arbitrary Taylor degree $k$, establish their asymptotic conditional bias and variance, and propose two approaches for estimating $a(x)$ and $b(x)$ when they are unknown. A simulation study across two structurally different ODEs shows that DE1-$k$ estimation with automatic degree selection reduces both estimation error and ODE-consistency error relative to unconstrained local linear regression. We then apply the framework to the firebrand burning-rate experiment of Albini (1979), modelling the density-loss curve of wind-driven firebrands with a physically motivated forced-convection ODE; the DE-constrained estimator outperforms local linear regression in the sparsest species-diameter groups, where physical structure is most valuable in compensating for scarce data. We further examine the robustness of the LQL model to misspecification against a local quasi-exponential alternative. Together, these results extend the DE-constrained regression paradigm to a broad class of physically motivated linear models and provide a practical estimation tool for fire science and other application areas where mechanistic knowledge is available but data are sparse.
\end{abstract}


\section{Introduction}

In recent years, British Columbia (B.C.) has faced increasingly severe wildfire seasons, driven by a combination of persistent drought, a warming climate, and the recurrence of overwintering ``zombie fires.'' By early July 2025, for example, 78 active wildfires were burning across the province according to the B.C. Wildfire Dashboard \citep{BCfire}, with a major blaze near Lytton prompting a local state of emergency and evacuation orders and forcing a temporary airspace closure at Kelowna International Airport. Episodes of this kind illustrate a broader trend: wildfire seasons across Canada and other fire-prone regions are growing longer, more intense, and more difficult to predict, which has renewed interest in quantitative tools for understanding how fires spread.

One of the most consequential mechanisms of fire spread is spotting: burning fragments of bark and wood, or firebrands, are lofted by the convective plume of a fire, carried downwind, and can ignite new fires well ahead of the main front. Predicting spotting distance requires understanding how quickly a firebrand loses mass and density while airborne, that is, its burning rate. Laboratory fire experiments play a crucial role in building this understanding under controlled conditions. In this paper, we study the firebrand burning-rate data from the wind-tunnel experiment of \citet{albini1979spot}, a foundational study in wildfire science that has informed spot-fire prediction models for over four decades.

Modelling the burning-rate curve is a nonparametric regression problem with a twist: physical reasoning about heat transfer and mass loss under forced convection suggests that the curve should approximately satisfy a first-order ordinary differential equation (ODE), even though the exact functional form of the curve is unknown and the data are sparse, with as few as three to five observations per experimental group. This is precisely the setting addressed by differential equation-constrained local polynomial regression (DE-constrained LPR), a nonparametric framework that embeds a known or partially known ODE into local polynomial fitting so that the physical structure of the problem, rather than the data alone, guides estimation where data are scarce. \citet{ge2026differential} introduced DE-constrained local polynomial (DE1-$k$) estimators for the exponential growth model $g'(x) = \lambda g(x)$ and showed that they substantially outperform unconstrained local polynomial regression in sparse-design settings. This complements a distinct strand of statistics that embeds ODEs into regression through globally parametric or semiparametric means, such as parameter cascading and generalized profiling \citep{ramsay2007parameter,brunel2008parameter,wang2022pcode}; unlike those approaches, DE-constrained LPR does not require solving the ODE explicitly or assuming a fully parametric error structure, which makes it attractive when the physical model is only approximately known or when a closed-form solution to the ODE is unavailable.

The exponential growth model captures ODEs of the restricted form $a(x) = \lambda$, $b(x) = 0$. Many physically motivated models, including the firebrand burning-rate model considered here, instead call for the general first-order linear ODE $g'(x) = a(x)g(x) + b(x)$, in which the rate of change of $g$ depends both on the current state $g(x)$, through $a(x)$, and on an external forcing term $b(x)$ that evolves independently of $g$. In this paper we extend the DE-constrained LPR framework to this general linear ODE structure, which we call the local quasi-linear (LQL) model.

The remainder of the paper is organized as follows. Section~\ref{sec:methodology} introduces the local quasi-linear regression model. Section~\ref{sec:estimation} derives the DE-constrained local polynomial (DE1-$k$) estimators and two approaches for estimating $a(x)$ and $b(x)$ when they are unknown. Section~\ref{sec:theory} establishes the asymptotic conditional bias and variance of the estimators. Section~\ref{sec:application} applies the framework to the Albini firebrand burning-rate data. Section~\ref{sec:misspecification} examines robustness of the LQL model to misspecification against a local quasi-exponential alternative. Section~\ref{sec:simulation} reports a simulation study comparing the DE-constrained estimator with unconstrained local linear regression across two additional ODE models. Section~\ref{sec:discussion} discusses the implications and limitations of the results, and Section~\ref{sec:conclusions} concludes.

\section{Methodology}
\label{sec:methodology}

\subsection{Differential Equation-Constrained Regression Model}

Consider the nonparametric regression model
    \begin{equation*}
     Y_i= g(X_i)+\varepsilon_i,  \quad i=1,2,\ldots,n,
     \end{equation*}
    where $g(\cdot)$ is an unknown smooth function and $\varepsilon_i$, $i=1,\ldots,n$, are independent random errors. When $X$ is a physically meaningful covariate, such as time, background scientific knowledge often implies that $g$ approximately satisfies a known or partially known ordinary differential equation (ODE). Embedding this information into the local polynomial fitting criterion, rather than discarding it, allows an estimator to borrow strength from the physical model precisely where data are sparse and unconstrained smoothers are least reliable. \citet{ge2026differential} developed this idea, which we refer to as differential equation-constrained local polynomial regression (DE-constrained LPR), for the exponential growth model $g'(x) = \lambda g(x)$ and demonstrated its advantage over unconstrained local polynomial regression under sparse designs. In this paper we generalize the framework to the general first-order linear ODE.

\subsection{Local Quasi-Linear Regression Model}

We consider the nonparametric regression model for a bivariate random vector $(X,Y)$,
    \begin{equation*}
     Y= g(X)+\sigma(X)\varepsilon,
     \end{equation*}
    where $\varepsilon$ is a random error with mean 0 and variance 1, independent of $X$. Here $g(x) = \mathrm{E}(Y \mid X=x)$ is the unknown regression function and $\sigma^2(x)=\mathrm{Var}(Y \mid X=x)$ is the conditional variance function. Suppose we know that
     \begin{equation}
     g'(x)=a(x)g(x)+b(x),
     \label{model:LQL}
    \end{equation}
\noindent that is, we have a differential equation-constrained model. We call this the \emph{local quasi-linear} (LQL) model, since the constraining ODE has a linear form. The exponential growth model of \citet{ge2026differential} is recovered as the special case $a(x)=\lambda$, $b(x)=0$. For the two functions $a(x)$ and $b(x)$, we consider three cases:

\begin{itemize}
\item[(i)] $a(x)$ and $b(x)$ are known;
\item[(ii)] $a(x)$ and $b(x)$ are unknown;
\item[(iii)] $a(x)$ and $b(x)$ are partly known, for example $a(x;\Theta)=\theta_1 x+\theta_2$ and $b(x;\Phi)=\phi_1 x+\phi_2$ with unknown $\theta_1,\theta_2,\phi_1,\phi_2$.
\end{itemize}

\section{The Proposed Estimations}
\label{sec:estimation}

Suppose we observe data $(x_i,y_i)$, $i=1,\ldots,n$, from model~(\ref{model:LQL}). We propose a DE-constrained approach to estimate the mean function $g(x)$ using the information encoded in the ODE, and we discuss how to estimate $a(x)$ and $b(x)$ when these are unknown.

\subsection{DE-Constrained Local Polynomial Estimation}
\label{sec:de-lpr}

Suppose first that $a(x)$ and $b(x)$ are known. For $x$ in a neighborhood of $x_i$, a first-order Taylor expansion of $g(x_i)$ combined with the ODE constraint $g'(x)=a(x)g(x)+b(x)$ gives the first-degree local approximant
\begin{equation*}
    g_1^*(x_i) = g(x) + (x_i-x)\{a(x)g(x)+b(x)\},
\end{equation*}
and a second-order expansion, using $g''(x)=(a'(x)+a^2(x))g(x)+a(x)b(x)+b'(x)$, gives the second-degree approximant
\begin{equation*}
    g_2^*(x_i) = g(x) + (x_i-x)g'(x) + \frac{(x_i-x)^2}{2}\left\{(a'(x)+a^2(x))g(x)+a(x)b(x)+b'(x)\right\}.
\end{equation*}
Higher-degree approximants follow from the recursive relationship
\begin{equation}
g^{(k+1)}(x) = \sum_{l=0}^{k}\binom{k}{l}a^{(l)}(x)g^{(k-l)}(x) + b^{(k)}(x).
\label{eqn:derivativeLinear}
\end{equation}

The degree-$k$ DE-constrained estimator $\hat{g}_k(x)$ minimizes the locally weighted least squares criterion
\begin{equation}
\sum_{i=1}^n \{y_i - g_k^*(x_i)\}^2 K_h(x_i-x)
\label{equ:genmethod}
\end{equation}
with respect to $g(x)$, where $K_h(\cdot)=K(\cdot/h)/h$ for kernel $K$ and bandwidth $h$. We refer to $\hat{g}_k(x)$ as the local DE1-$k$ estimator, where the first `1' denotes the order of the constraining ODE and $k$ denotes the degree of the Taylor expansion. Because~(\ref{equ:genmethod}) is quadratic in $g(x)$, the minimizer is available in closed form for every $k$; no iterative solver is required, and the estimator is correspondingly stable, failing only when the local design matrix is singular, which occurs when fewer than $k+1$ distinct design points fall within the bandwidth.

Minimizing~(\ref{equ:genmethod}) at $k=1$ gives the \emph{DE1-1} estimator
\begin{align}
\widehat{g}_1(x)
= \frac{\sum_{i=1}^n (y_i-(x_i-x)b(x))(1+(x_i-x)a(x))K_h(x_i-x) }{\sum_{i=1}^n (1+(x_i-x)a(x))^2K_h(x_i-x)}.
\label{equ:wlslinear1}
\end{align}
Denoting $\phi(x,x_i-x) = 1+(x_i-x)a(x)+\frac{1}{2}(x_i-x)^2(a'(x)+a^2(x))$, minimizing~(\ref{equ:genmethod}) at $k=2$ gives the \emph{DE1-2} estimator
\begin{small}
\begin{align}
\widehat{g}_2(x)
= \frac{
    \displaystyle\sum_{i=1}^n
    \Bigl(y_i - (x_i-x)b(x)
          - \tfrac{1}{2}(x_i-x)^2\bigl(a(x)b(x)+b'(x)\bigr)
    \Bigr)\,\phi(x,x_i-x)\, K_h(x_i-x)
  }{
    \displaystyle\sum_{i=1}^n \phi^2(x,x_i-x)\, K_h(x_i-x)
  }.
\label{equ:wlslinear2}
\end{align}
\end{small}
The expression for the $k$th-degree estimator with $k>2$ follows analogously, using~(\ref{eqn:derivativeLinear}) to construct $g_k^*(x_i)$.

\subsection{Estimation of $a(x)$ and $b(x)$}
\label{sec:estimate-ab}

When $a(x)$ and $b(x)$ are unknown, we propose two approaches to estimate them.

\bigbreak
\textbf{(1) Pilot estimator with weighted linear regression.}
\bigbreak

\noindent Step 1: Use the local constant estimator as a pilot estimator for $g(x)$, denoted $\widehat{g}_0=\widehat{g}_{LC}(x)$.

\noindent Step 2: Substitute $\widehat{g}_0$ into the locally weighted least squares criterion and estimate $a(x)$ and $b(x)$ by weighted linear regression with design matrix
$$
\textbf{X}=
\begin{pmatrix}
(x_1-x)\widehat{g}_0(x_1) & (x_1-x)  \\
(x_2-x)\widehat{g}_0(x_2) & (x_2-x)  \\
\vdots & \vdots \\
(x_n-x)\widehat{g}_0(x_n) & (x_n-x)
\end{pmatrix}
$$
and response vector
$$
Y=
\begin{pmatrix}
y_1-\widehat{g}_0(x_1)   \\
y_2-\widehat{g}_0(x_2)  \\
\vdots  \\
y_n-\widehat{g}_0(x_n)
\end{pmatrix}.
$$

\noindent Step 3: Substitute $\widehat{a}(x)$ and $\widehat{b}(x)$ into~(\ref{equ:wlslinear1}) to obtain the DE-constrained estimator $\widehat{g}_1(x)$.

\bigbreak
\textbf{(2) Direct locally weighted least squares.}
\bigbreak

\noindent Step 1: Rewrite the locally weighted least squares criterion as
\begin{equation}
\sum_{i=1}^n \{y_i-g(x)-(x_i-x)a(x)g(x)-(x_i-x)b(x)\}^2 K_h(x_i-x).
\label{equ:wls1_1}
\end{equation}

\noindent Step 2: Denote $g_0=g(x)$ and $\delta_0=a(x)g(x)+b(x)$, and rewrite the model as $y_i \approx g_0+(x_i-x)\delta_0$. Obtain $\hat{g}_0$ and $\hat\delta_0$ by weighted linear regression with design matrix
$$
\mathbf{X}=
\begin{pmatrix}
1 & (x_1-x)  \\
1 & (x_2-x)  \\
\vdots & \vdots \\
1 & (x_n-x)
\end{pmatrix}
$$
and response vector $Y=(y_1,\ldots,y_n)^{\top}$.

\noindent Step 3: The first-degree DE-constrained estimate is $\widehat{g}_1(x)=\hat{g}_0$. Since $\hat\delta_0$ estimates the combination $a(x)g(x)+b(x)$ rather than $a(x)$ and $b(x)$ individually, the two functions cannot be recovered separately from this step alone without additional information; in practice, $\hat\delta_0$ can be substituted directly for $a(x)\hat{g}_0+b(x)$ in the DE-constrained estimator~(\ref{equ:wlslinear1}).

Related bias-reduction techniques for nonparametric regression under model uncertainty, such as the approach of \citet{cheng2018bias}, suggest a further extension in which the DE constraint is combined with a variance-function-based correction; we do not pursue this combination here, but return to the closely related question of robustness under model misspecification in Section~\ref{sec:misspecification}.

\section{Theoretical Properties}
\label{sec:theory}

In this section we establish the asymptotic conditional bias and variance of the DE1-$k$ estimator $\widehat{g}_k(x)$ of Section~\ref{sec:de-lpr}, for general degree $k$, under the following assumptions on model~(\ref{model:LQL}):

\begin{itemize}
  \item[(I)] $g(x)$, the mean function, has a bounded and continuous $(k+2)$th derivative in a neighborhood of $x$.
  \item[(II)] $f(x)$, the design density with support $[a,b]$, is twice continuously differentiable and positive.
  \item[(III)] $K(\cdot)$ is a nonnegative, symmetric, bounded probability density function with $\int K(w)\,dw=1$, finite moments up to sixth order, and $R(K)=\int K^2(w)\,dw<\infty$. We write $\mu_j=\int w^jK(w)\,dw$. Kernels with compact support (e.g., Epanechnikov) and kernels with rapidly decaying tails (e.g., Gaussian) both satisfy this condition.
  \item[(IV)] The conditional variance $\sigma^2(x)$ is a smooth function on $[a,b]$.
\end{itemize}

\begin{theorem}[Asymptotic Conditional Bias]
\label{thm:bias}
For model~(\ref{model:LQL}), under assumptions~\textnormal{(I)--(III)}, with $x\in(a+h,b-h)$, the DE1-$k$ estimator $\hat{g}_k(x)$ has asymptotic conditional bias
\begin{equation}
\mathrm{Bias}(\widehat{g}_k(x)\mid x_1,\ldots,x_n) =
\frac{1}{(k+1)!}g^{(k+1)}(x)h^{k+1}\mu_{k+1}+o_p(h^{k+1}), \quad k \text{ odd},
\end{equation}
and for even $k$,
\begin{equation}
\mathrm{Bias}(\widehat{g}_k(x)\mid x_1,\ldots,x_n) =
\left(\frac{g^{(k+2)}(x)}{(k+2)!}+\frac{g^{(k+1)}(x)}{(k+1)!}\frac{f'(x)}{f(x)}\right)h^{k+2}\mu_{k+2}+o_p(h^{k+2}),
\end{equation}
where $g^{(k+1)}(x)$ and $g^{(k+2)}(x)$ are computed from the recursion~(\ref{eqn:derivativeLinear}). A derivation for the case $k=1$, obtained via the conditional expectation of $\widehat{g}_1(x)$ and the mean value theorem for integrals, is given in Appendix~\ref{app:proofs}.
\end{theorem}

\begin{remark}
We illustrate the bias formulae for $k=1$ and $k=2$ using~(\ref{eqn:derivativeLinear}) with $g'(x)=a(x)g(x)+b(x)$.

\medskip
\noindent\textbf{Example ($k=1$, odd).} A single application of~(\ref{eqn:derivativeLinear}) gives
\begin{equation*}
    g''(x) = \bigl(a'(x)+a^2(x)\bigr)g(x)+a(x)b(x)+b'(x),
\end{equation*}
so that
\begin{equation}
    \mathrm{Bias}\bigl(\widehat{g}_1(x)\mid x_1,\ldots,x_n\bigr)
    = \frac{h^2\mu_2}{2}\Bigl[\bigl(a'(x)+a^2(x)\bigr)g(x)+a(x)b(x)+b'(x)\Bigr]+o_p(h^2).
    \label{eqn:bias_k1_ex}
\end{equation}

\medskip
\noindent\textbf{Example ($k=2$, even).} A second application of~(\ref{eqn:derivativeLinear}) gives
\begin{equation*}
    g'''(x) = \bigl(a''(x)+3a(x)a'(x)+a^3(x)\bigr)g(x)+\bigl(a'(x)+a^2(x)\bigr)b(x)+a(x)b'(x)+b''(x),
\end{equation*}
so that
\begin{multline}
    \mathrm{Bias}\bigl(\widehat{g}_2(x)\mid x_1,\ldots,x_n\bigr)
    = \frac{h^3\mu_3}{6}\Bigl[\bigl(a''+3aa'+a^3\bigr)g+\bigl(2a'+a^2\bigr)b+ab'+b''\Bigr] \\
    +\, \frac{h^3\mu_3}{2}\Bigl[\bigl(a'+a^2\bigr)g+ab+b'\Bigr]\frac{f'(x)}{f(x)}
    + o_p(h^3),
    \label{eqn:bias_k2_ex}
\end{multline}
where all coefficient functions are evaluated at $x$. For symmetric kernels, $\mu_3=0$ and the leading term of $\mathrm{Bias}(\widehat{g}_2)$ vanishes, so the bias is $o_p(h^3)$, illustrating the bias reduction achieved by even-degree DE1-$k$ estimators relative to odd-degree ones.
\end{remark}

\begin{theorem}[Asymptotic Conditional Variance]
\label{thm:variance}
Under assumptions~\textnormal{(I)--(IV)}, with $x\in(a+h,b-h)$, the DE1-$k$ estimator $\hat{g}_k(x)$ has asymptotic conditional variance
\begin{equation}
\mathrm{Var}(\widehat{g}_k(x)\mid x_1,\ldots,x_n) = \frac{\sigma^2(x) R(K)}{nhf(x)}+o_p\!\left(\frac{1}{nh}\right),
\end{equation}
which does not depend on $k$ to this order. A derivation for $k=1$ is given in Appendix~\ref{app:proofs}.
\end{theorem}

Theorems~\ref{thm:bias} and~\ref{thm:variance} together show that the DE-constrained local polynomial regression approach reduces asymptotic bias, at a rate governed by the degree $k$ of the Taylor expansion embedded in the ODE constraint, without inflating the asymptotic variance relative to unconstrained local polynomial regression of the same order. This bias--variance trade-off is the mechanism by which DE-constrained estimation improves on unconstrained local polynomial regression in the sparse-data settings examined in Sections~\ref{sec:application} and~\ref{sec:simulation}.

\section{Application}
\label{sec:application}

In this section we apply the DE-constrained regression approach of Sections~\ref{sec:estimation} and~\ref{sec:theory} to the firebrand burning-rate data of \citet{albini1979spot}. When a tree burns, it generates cylindrical wooden firebrands, fragments of branches and twigs, that are lofted by the fire plume, transported downwind, and can ignite spot fires far from the main fire front. A key quantity governing how far a firebrand can travel is its burning rate: how quickly it loses mass and density while airborne. \citet{albini1979spot} conducted controlled laboratory experiments in which cylindrical firebrands from four wood species were burned under cross-flow wind conditions at different wind speeds and initial sizes, with final density recorded at different burn times. Our goal is to estimate, for each species, the density-ratio function $g(t)$, that is, how the ratio of final to initial density evolves over time, using a nonparametric method constrained by a physically motivated ODE. Since time is the primary covariate in this application, we write the mean function as $g(t)$ rather than $g(x)$.

\subsection{Albini Burning Rate Data}

The data contain $n=33$ observations, each recording a single firebrand experiment stopped at a fixed burn time. The first six observations are shown in Table~\ref{table:burningdata}.

  \begin{table}[ht]
\centering
\scalebox{0.8}{
\begin{tabular}{cccccccccccc}
\hline
 & species & diameterI & massI & densityI & massF & densityF & $DD_o$ & $rDrD_o$ & wind & time & length \\
\hline
1 & PP & 1 & 53.36 & 0.610 & 4.16 & 0.240 & 0.445 & 0.175 & 15 & 360 & 5 \\
2 & PP & 1 & 47.20 & 0.520 & 6.47 & 0.260 & 0.523 & 0.262 & 15 & 300 & 5 \\
3 & PP & 1 & 44.59 & 0.584 & 5.32 & 0.220 & 0.562 & 0.212 & 15 & 240 & 5 \\
4 & PP & 1 & 31.33 & 0.472 & 9.42 & 0.239 & 0.771 & 0.390 & 15 & 180 & 5 \\
5 & PP & 1 & 41.33 & 0.559 & 13.31 & 0.467 & 0.621 & 0.519 & 15 & 120 & 5 \\
6 & ES & 1 & 48.44 & 0.706 & 6.99 & 0.329 & 0.557 & 0.259 & 15 & 360 & 5 \\
\hline
\end{tabular}
}
\caption{The first six observations in the Albini (1979) firebrand burning rate data.}
\label{table:burningdata}
\end{table}

The objects were oven-dry limbwood samples of four species: PP (Ponderosa pine), ES (Engelmann spruce), WL (Western larch), and WRC (Western red cedar). The variables are: \texttt{species}; \texttt{diameterI}, initial firebrand diameter (in.); \texttt{massI}/\texttt{massF}, initial/final mass (g); \texttt{densityI}/\texttt{densityF}, initial/final density (g/cm$^3$); $DD_o$, final/initial diameter ratio; $rDrD_o$, dimensionless burning rate; \texttt{wind}, wind speed (mph); \texttt{time}, burn duration (s); and \texttt{length}, firebrand length (in.). The data are structured in eight groups (four species $\times$ two initial diameters, 0.5 and 1.0 in.), with three to five observations per group.

\subsection{State Variable, ODE, and DE-Constrained Model}

The state variable is the dimensionless density ratio $g(t)=\rho_F(t)/\rho_I$, where $\rho_F(t)$ is the density at burn time $t$ and $\rho_I$ is the initial density; by definition $g(0)=1$ for every experiment. Since $\rho_I$ varies across experiments (0.35--0.79 g/cm$^3$), normalizing by it places all observations on a common scale with $g(t)\in(0,1]$.

The physical model for density loss under forced convection is
\begin{equation}
g'(t) = \beta u \left(\frac{e^{\lambda t}}{\rho_I d_I} - 2\lambda g(t)\right),
\label{eqn:AlbiniODE}
\end{equation}
where $u$ is wind speed, $\rho_I$ and $d_I$ are the initial density and diameter (fixed within each experiment), and $\beta>0$, $\lambda>0$ are unknown species-specific parameters, with $\lambda$ controlling the timescale of combustion and $\beta$ controlling how efficiently wind speed drives density loss. In standard linear form, $g'(t)+2\beta u\lambda\, g(t) = (\beta u/(\rho_I d_I))e^{\lambda t}$, so that~(\ref{eqn:AlbiniODE}) is an instance of the LQL model~(\ref{model:LQL}) with $a(t)=-2\lambda\beta u$ and $b(t)=\beta u\,e^{\lambda t}/(\rho_I d_I)$: a case where $a(t)$ and $b(t)$ are partly known, in the sense of case (iii) of Section~\ref{sec:methodology}, since their functional form is known but $\beta$ and $\lambda$ must be estimated. Incorporating~(\ref{eqn:AlbiniODE}) and the initial condition into the regression framework gives
\begin{equation}
y_i = g(t_i) + \varepsilon_i, \quad
g'(t) = \beta u\!\left(\frac{e^{\lambda t}}{\rho_I d_I} - 2\lambda g(t)\right), \quad
g(0) = 1, \quad i = 1,\ldots,n,
\label{model:Albinilinear}
\end{equation}
where $y_i=\rho_{F,i}/\rho_{I,i}$ and the $\varepsilon_i$ are independent with mean zero and variance $\sigma^2$.

\subsection{Estimation}
\label{sec:albini_estimation}

Estimation of model~(\ref{model:Albinilinear}) proceeds in two stages. In Stage~1, pilot estimates $\hat\beta$ and $\hat\lambda$ are obtained by nonlinear least squares (NLS), separately for each species. Multiplying~(\ref{eqn:AlbiniODE}) by $\rho_I$ gives the ODE for $\rho_F(t)=\rho_I g(t)$,
\begin{equation}
\rho_F'(t) + 2\lambda\beta u\,\rho_F(t) = \frac{\beta u}{d_I}\,e^{\lambda t}, \qquad \rho_F(0) = \rho_I,
\label{eqn:rhoF_ODE}
\end{equation}
which, solved by integrating factor $\mu(t)=e^{2\lambda\beta u\cdot t}$, yields the explicit solution
\begin{equation}
g(t) = A\,e^{\lambda t} + (1-A)\,e^{-2\lambda\beta u\cdot t}, \qquad g(0)=1, \qquad A = \frac{\beta u}{\lambda\rho_I d_I(1+2\beta u)}.
\label{eqn:g_explicit}
\end{equation}
This is used as the Stage~1 NLS response model, fitted separately per species; the resulting pilot estimates $(\hat\beta,\hat\lambda)$ are reported in Table~\ref{tab:stage1}.

\begin{table}[ht]
\centering
\small
\setlength{\tabcolsep}{4pt}
\begin{tabular}{lllllr}
\hline
Species & $\hat{\beta}$ & $\hat{\lambda}$ & $\beta$ & $\lambda$ & Res.\ SE \\
\hline
ES  & $6.43\times10^{-5}$ & $4.256\times10^{-3}$ & $^{*}$($p=0.018$) & $^{**}$($p=0.006$) & 0.157 \\
PP  & $4.53\times10^{-5}$ & $4.870\times10^{-3}$ & $^{*}$($p=0.010$) & $^{**}$($p=0.004$) & 0.112 \\
WRC & $2.32\times10^{-5}$ & $4.555\times10^{-3}$ & ns($p=0.213$)     & $^{**}$($p=0.004$) & 0.076 \\
WL  & $4.40\times10^{-6}$ & $3.519\times10^{-3}$ & ns($p=0.800$)     & $^{**}$($p=0.001$) & 0.037 \\
\hline
\end{tabular}
\caption{Stage-1 NLS pilot estimates by species. $^{**}p<0.01$, $^{*}p<0.05$, ns: not significant. ES = Engelmann Spruce, PP = Ponderosa Pine, WRC = Western Red Cedar, WL = Western Larch.}
\label{tab:stage1}
\end{table}

With $\hat\beta$ and $\hat\lambda$ fixed, Stage~2 estimates $g(t_0)$ nonparametrically using the DE-constrained local polynomial framework of Section~\ref{sec:de-lpr}, with $a(t)$ and $b(t)$ as identified above. The required ODE derivatives at $t_0$, obtained from~(\ref{eqn:derivativeLinear}), are
\begin{align}
g'(t_0)   &= \frac{\beta u}{\rho_I d_I}e^{\lambda t_0} - 2\lambda\beta u\,g_0,
\label{eqn:g1}\\[4pt]
g''(t_0)  &= \frac{\lambda\beta u(1-2\beta u)}{\rho_I d_I}e^{\lambda t_0} + 4\lambda^2\beta^2 u^2\,g_0,
\label{eqn:g2}\\[4pt]
g'''(t_0) &= \frac{\lambda^2\beta u(1-2\beta u+4\beta^2u^2)}{\rho_I d_I}e^{\lambda t_0} - 8\lambda^3\beta^3 u^3\,g_0,
\label{eqn:g3}
\end{align}
where $g_0=g(t_0)$. The DE1-1, DE1-2, and DE1-3 estimators correspond to truncating the Taylor expansion at degree $k=1,2,3$ respectively, using a Gaussian kernel with $h=50$ s (widened to $1.25h=62.5$ s for DE1-3), and minimizing the resulting weighted least squares objective over a grid of 401 equally spaced evaluation points by the Gauss--Newton algorithm. Results are compared against unconstrained local linear regression (degree-1 \texttt{locpoly}) and the explicit pilot solution curve~(\ref{eqn:g_explicit}).

\subsection{Data Analysis}
\label{sec:albini_data_analysis}

\begin{figure}[ht]
\centering
\includegraphics[width=\textwidth]{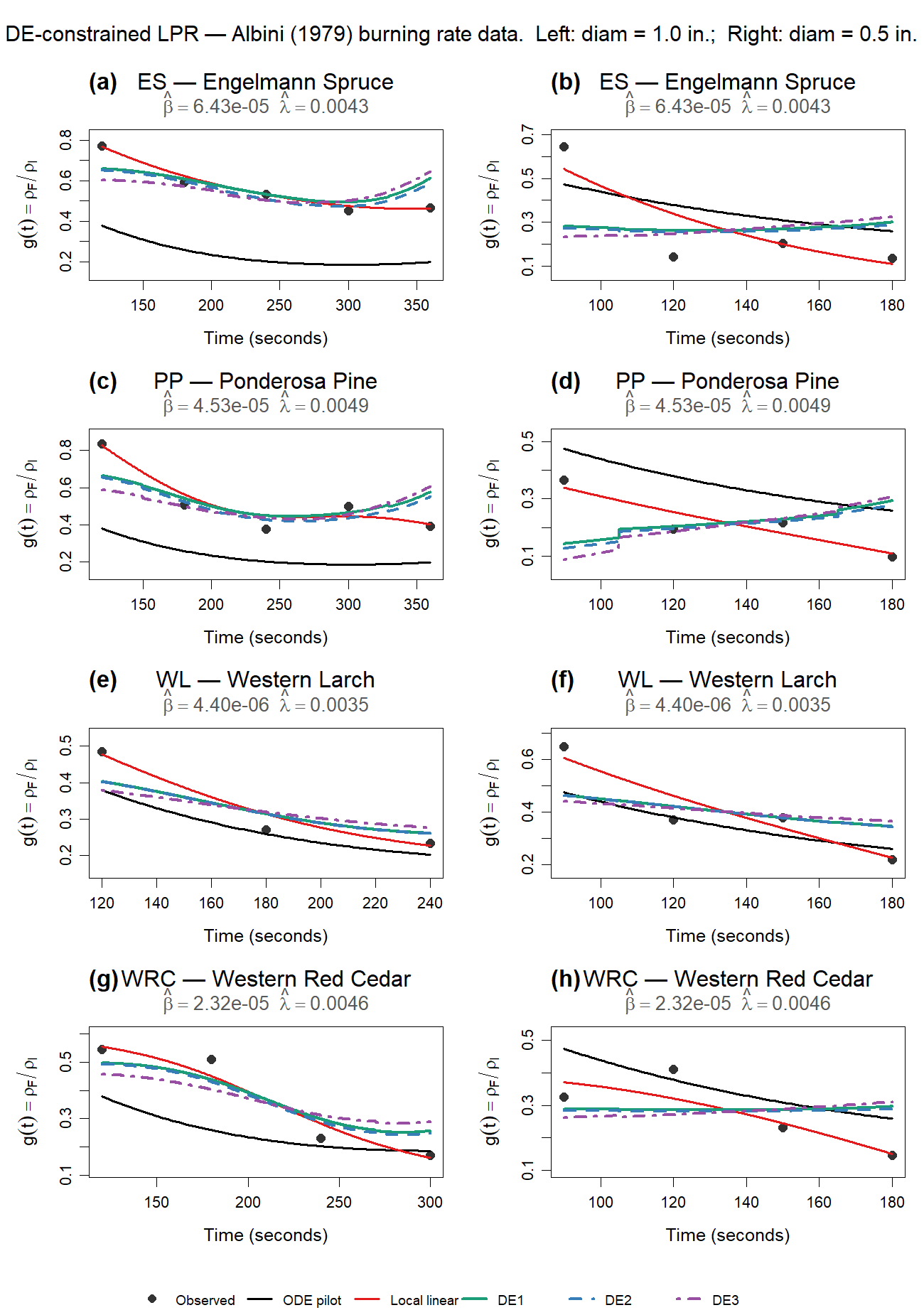}
\caption{DE-constrained LPR fitted to the Albini (1979) burning rate data. State variable $g(t)=\rho_F(t)/\rho_I$. Left column: $d_I=1.0$ in.; right column: $d_I=0.5$ in. Blue curves: DE1-1 (dashed), DE1-2 (dotted), DE1-3 (dot-dash); red: local linear; black: ODE pilot solution~(\ref{eqn:g_explicit}). Gaussian kernel, $h=50$ s.}
\label{fig:albini_fits}
\end{figure}

Figure~\ref{fig:albini_fits} shows the fitted curves for all eight species-diameter groups, and Table~\ref{tab:loocv} reports leave-one-out cross-validation (LOO-CV) mean squared error for each method.

The parameter $\lambda$ is significant for all four species ($p<0.01$), confirming systematic density loss in every case. The parameter $\beta$ is significant only for ES and PP ($p<0.05$), indicating that wind speed is a detectable driver of burning for these species; for WRC and WL, burning is governed primarily by $\lambda$ alone.

\begin{table}[ht]
\centering
\begin{tabular}{llrllll}
\hline
Species & Diam (in.) & $n$ & LL & DE1-1 & DE1-2 & DE1-3 \\
\hline
ES  & 1.0 & 5 & \textbf{0.0060} & 0.0524 & 0.0457 & 0.0592 \\
ES  & 0.5 & 4 & \textbf{0.0904} & 0.1238 & 0.1245 & 0.1343 \\
PP  & 1.0 & 5 & \textbf{0.0273} & 0.1298 & 0.1159 & 0.1326 \\
PP  & 0.5 & 4 & \textbf{0.0080} & 0.0431 & 0.0425 & 0.0489 \\
WL  & 1.0 & 3 & 0.0238 & \textbf{0.0227} & 0.0227 & 0.0249 \\
WL  & 0.5 & 4 & \textbf{0.0169} & 0.0347 & 0.0347 & 0.0377 \\
WRC & 1.0 & 4 & 0.0300 & 0.0324 & \textbf{0.0296} & 0.0437 \\
WRC & 0.5 & 4 & \textbf{0.0170} & 0.0268 & 0.0258 & 0.0300 \\
\hline
\end{tabular}
\caption{LOO-CV MSE by group. Bold: best estimator per group. LL = local linear.}
\label{tab:loocv}
\end{table}

The DE-constrained estimators outperform local linear regression in the two sparsest groups, WL with $d_I=1.0$ in. ($n=3$, DE1-1 wins) and WRC with $d_I=1.0$ in. ($n=4$, DE1-2 wins). Local linear regression performs better in the remaining six groups, where $n\geq4$ and the pilot estimation error in $\hat\beta,\hat\lambda$ outweighs the benefit of the ODE constraint; DE1-3 does not improve on DE1-2 in any group. These results confirm that DE-constrained LPR reduces prediction error relative to unconstrained local linear regression specifically in the sparsest data groups, consistent with the bias--variance trade-off established in Theorems~\ref{thm:bias} and~\ref{thm:variance}, while the advantage diminishes as group size increases, reflecting the trade-off between the structural information provided by the ODE constraint and the pilot parameter estimation error. Among the species studied, ES and PP exhibit significant wind-driven combustion, while WL and WRC do not, suggesting greater structural resistance to forced convection in the latter two species.

\section{Model Misspecification}
\label{sec:misspecification}

A model is misspecified when the constraint imposed on the estimator does not accurately reflect the true data-generating process. Statistical tools for detecting and diagnosing model misspecification include residual analysis, cross-validation, information criteria, and specification tests such as the Ramsey RESET test \citep{ramsey1969tests}; \citet{cheng2018bias} developed bias-reduction methods for nonparametric and semiparametric regression that are particularly useful when the assumed parametric model is misspecified. Here we examine, informally, how the LQL estimator of Section~\ref{sec:estimation} behaves when the assumed linear ODE constraint does not hold exactly.

Suppose the true data-generating process is not the LQL model~(\ref{model:LQL}) but instead a \emph{local quasi-exponential} (LQE) model,
\begin{equation}
y_i = g(x_i)e^{\varepsilon_i}, \quad i=1,\ldots,n,
\label{eqn:LQEdata}
\end{equation}
where $x_i$ and $\varepsilon_i$ are independent, $\varepsilon_i$ are uncorrelated errors with mean 0 and variance $\sigma^2$, and the mean function satisfies
\begin{equation}
g'(x)=\lambda g^{\alpha}(x)
\label{equ:LQE}
\end{equation}
for known parameters $\lambda$ and $\alpha$. When $\alpha=1$, model~(\ref{equ:LQE}) reduces to the exponential growth model $g'(x)=\lambda g(x)$, a special case of the LQL model with $a(x)=\lambda$, $b(x)=0$; for $\alpha\neq1$, however, the true rate of change of $g$ depends on $g(x)$ nonlinearly, and no constant or low-order polynomial pair $(a(x),b(x))$ reproduces~(\ref{equ:LQE}) exactly. If, unaware of this, we apply the LQL estimator~(\ref{equ:wlslinear1}) with $a(x)$ and $b(x)$ fixed at values appropriate to the linear case (for example, the local linearization $a(x_0)=\lambda\alpha g^{\alpha-1}(x_0)$, $b(x_0)=\lambda(1-\alpha)g^{\alpha}(x_0)$ of~(\ref{equ:LQE}) around a fixed working point $x_0$), the constraint is exact only at $x=x_0$ and drifts increasingly out of alignment with the true local slope as $|x-x_0|$ grows or as $\alpha$ departs from 1.

The practical consequence parallels the exponential-model case: the DE-constrained estimator's asymptotic bias in Theorem~\ref{thm:bias} is derived under the assumption that the constraint $g'(x)=a(x)g(x)+b(x)$ holds exactly, so that the only remaining bias is the usual $O(h^{k+1})$ (or $O(h^{k+2})$) smoothing bias, which vanishes as $h\to0$. Under misspecification, the estimator instead targets the constrained approximant $g_k^*(x_i)$ built from the wrong $(a,b)$ pair, and the resulting bias acquires an additional model-error term reflecting the discrepancy between the assumed and true local dynamics; unlike the smoothing bias, this term does not vanish as $h\to0$ or $n\to\infty$, and it grows with the degree of curvature mismatch $|\alpha-1|$ and with distance from $x_0$. This is the standard cost of constrained estimation: a constraint that is correct reduces bias relative to an unconstrained estimator at the same bandwidth, but a constraint that is wrong introduces a persistent bias that constrained estimation cannot average away. In practice, this argues for checking the plausibility of the assumed ODE, for example via the residual-based diagnostics referenced above, before relying on the DE-constrained estimator, and for treating the estimation approaches of Section~\ref{sec:estimate-ab}, which re-estimate $a(x)$ and $b(x)$ locally from the data rather than fixing them a priori, as a partial safeguard against this form of misspecification.

\section{Simulation}
\label{sec:simulation}

We conduct a simulation study to assess the performance of the DE1-$k$ estimator relative to unconstrained local linear regression across two examples of the general linear ODE model~(\ref{model:LQL}), each under two sample sizes. The simulation code was originally developed by the second author and subsequently adapted by the first author.

\subsection{Simulation Design}

\noindent\textbf{Example 1 (toy model).} The true function is
\begin{equation}
    g(x) = e^{-\cos(x)+1} - 1, \qquad x \in [0, 10],
    \label{eqn:sim_toy_g}
\end{equation}
which satisfies $g'(x) = a(x)g(x) + b(x)$ with $a(x) = b(x) = \sin(x)$. Observations are generated as $y_i = g(x_i) + \varepsilon_i$ with $\varepsilon_i \overset{\text{iid}}{\sim} N(0, 0.25)$, under a \emph{fixed} design with $n = 25$ equally spaced points on $[0, 10]$ and a \emph{random} design with $n = 100$ points drawn independently from $\mathrm{Uniform}(0, 10)$.

\medskip
\noindent\textbf{Example 2 (Michaelis--Menten model).} The true function is
\begin{equation}
    g(x) = \frac{V_{\max} x}{K_m + x}, \qquad x \in [0.2, 1.1],
    \label{eqn:sim_mm_g}
\end{equation}
with $V_{\max} = 191$ and $K_m = 0.06$, corresponding to the Puromycin enzyme kinetics data. This function satisfies $g'(x) = a(x)g(x)$ with $a(x) = K_m/(x(K_m+x))$ and $b(x) = 0$, so the constraining ODE is homogeneous linear. Observations are generated with $\varepsilon_i \overset{\text{iid}}{\sim} N(0, 225)$, under the same fixed ($n=25$) and random ($n=100$) designs, rescaled to $[0.2,1.1]$.

Each simulation is replicated $N_{\text{sim}} = 1000$ times.

\subsection{Methods and Evaluation Criteria}

Two estimators are compared in each replication: the \emph{local linear} (LL) estimator, the standard degree-1 local polynomial fit using \texttt{locpoly} from the \texttt{KernSmooth} package \citep{kernsmooth} with bandwidth from the direct plug-in rule \texttt{dpill}; and the \emph{DE-constrained} (DE) estimator, the DE1-$k$ estimator of Section~\ref{sec:de-lpr}, fitted for $k = 1,\ldots,8$ with bandwidth $h_k = \hat{h}\cdot k^{1/4}$, where $\hat{h}$ is the plug-in bandwidth from the pilot local linear fit. The best degree is selected as $k^* = \arg\min_{k\in\{1,\ldots,8\}}\mathrm{MAE}_k$, the degree minimizing the mean absolute ODE residual error defined below. Both estimators are evaluated on the same grid of $M=401$ equally spaced points over $[a,b]$.

Performance is measured by the mean absolute deviation error (MADE), $\mathrm{MADE} = M^{-1}\sum_{j=1}^{M}|\hat{g}(x_j)-g(x_j)|$, which measures how accurately the estimator recovers $g$, and the mean absolute derivative error (MAE),
\begin{equation}
    \mathrm{MAE} = \frac{1}{M-1}\sum_{j=1}^{M-1}
    \left|\frac{\hat{g}(x_{j+1})-\hat{g}(x_j)}{x_{j+1}-x_j} - \bigl(a(x_j)\hat{g}(x_j)+b(x_j)\bigr)\right|,
    \label{eqn:MAE}
\end{equation}
which measures how well the fitted curve satisfies the ODE constraint, with the derivative of $\hat{g}$ approximated by finite differences on the grid. Smaller values of both criteria indicate better performance.

\subsection{Results}

\begin{figure}[ht]
    \centering
    \includegraphics[width=\textwidth]{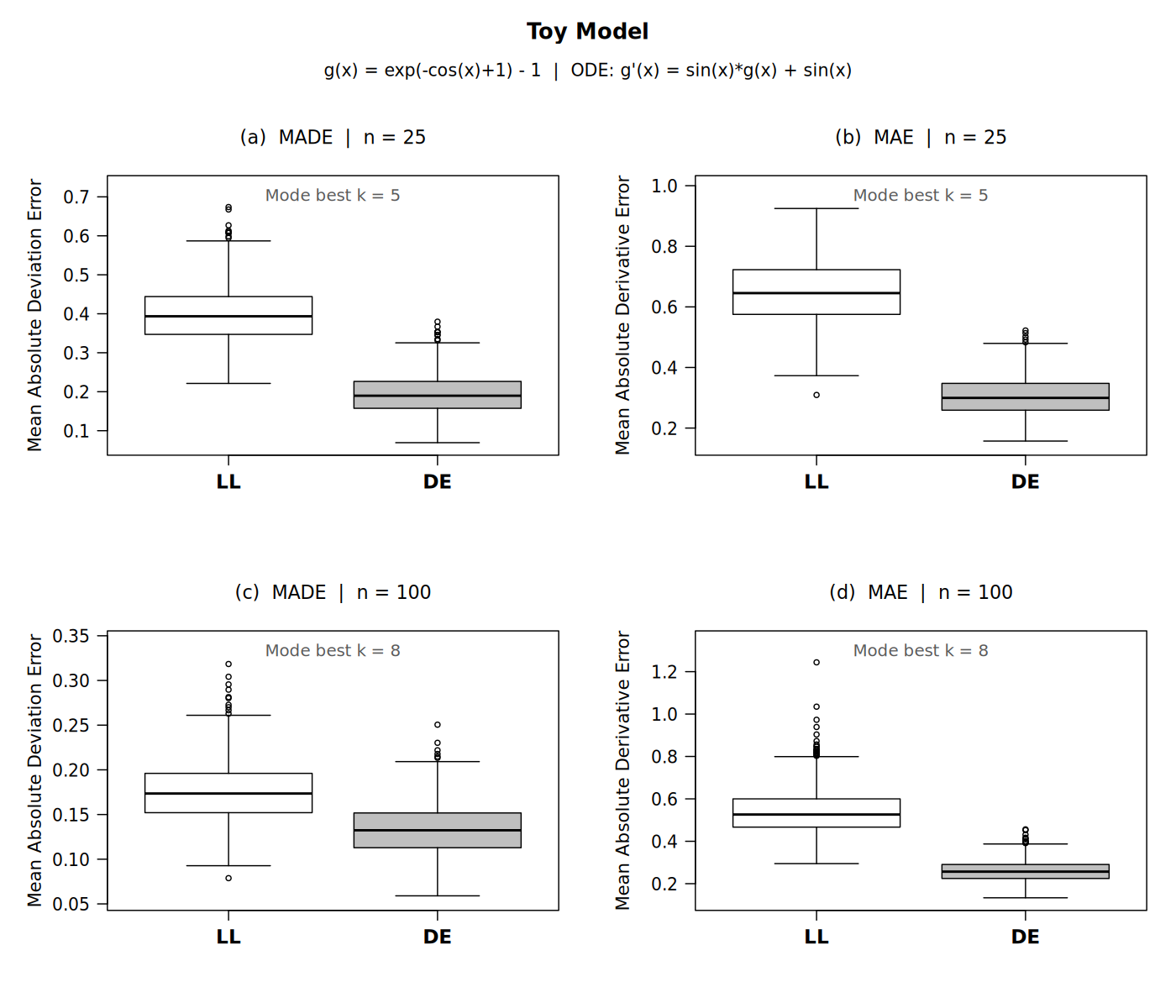}
    \caption{Simulation results for Example 1 (toy model). Each panel shows boxplots of MADE (left column) and MAE (right column) over 1000 replications, for local linear regression (LL) and the best DE-constrained estimator (DE). Upper row: $n=25$ fixed design; lower row: $n=100$ random design.}
    \label{fig:sim_toy}
\end{figure}

\begin{figure}[ht]
    \centering
    \includegraphics[width=\textwidth]{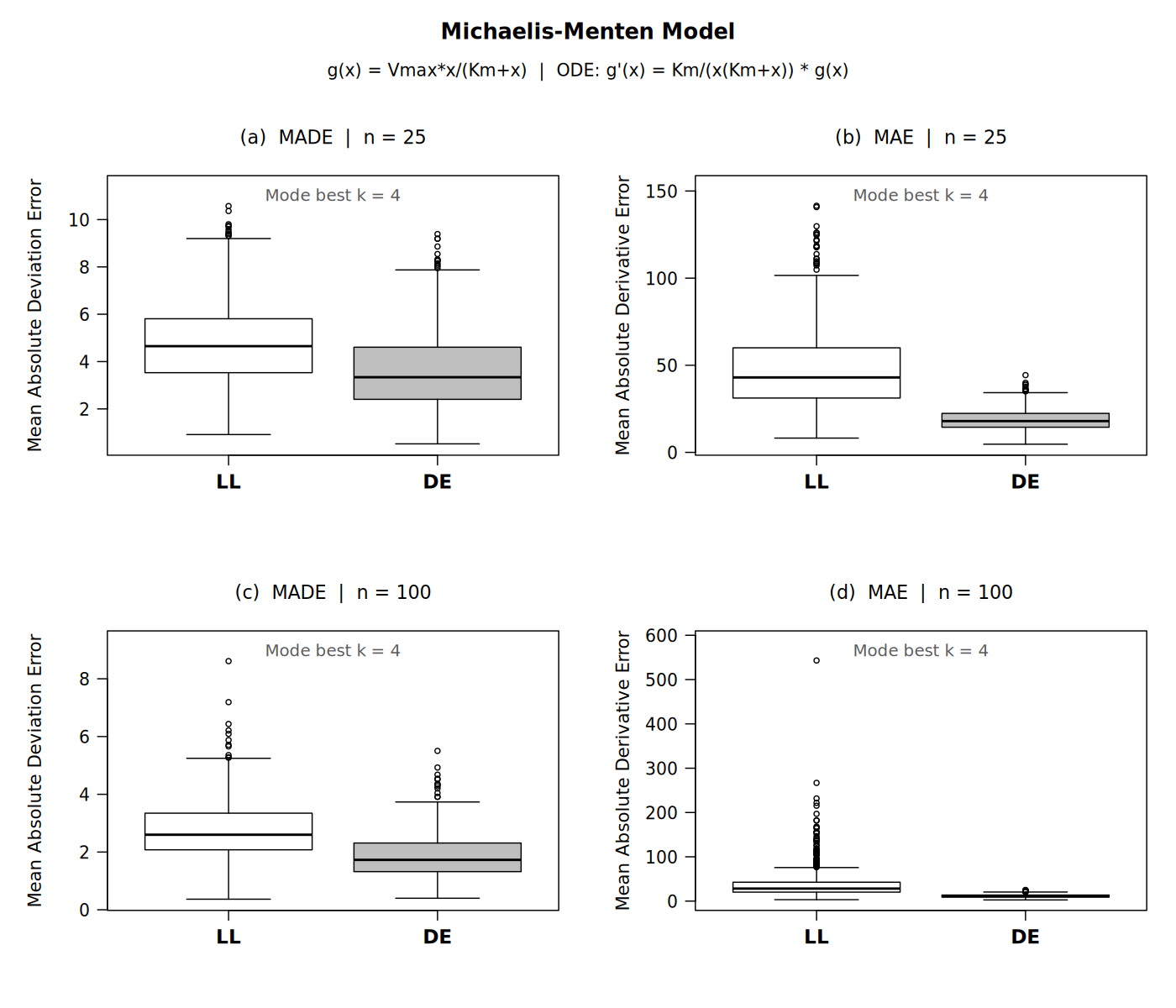}
    \caption{Simulation results for Example 2 (Michaelis--Menten model). Layout as in Figure~\ref{fig:sim_toy}.}
    \label{fig:sim_mm}
\end{figure}

Figures~\ref{fig:sim_toy} and~\ref{fig:sim_mm} display boxplots of MADE and MAE over 1000 replications for the toy and Michaelis--Menten examples, respectively.

In both examples and at both sample sizes, the DE estimator achieves lower MADE and MAE than local linear regression. The MAE improvement is consistently greater than the MADE improvement, particularly for the Michaelis--Menten model, reflecting the stronger ODE constraint imposed by the homogeneous structure $b(x)=0$. The selected degree $k^*$ differs between examples: for Example 1, the mode of $k^*$ is 5 at $n=25$ and 8 at $n=100$, indicating that the oscillatory nature of $g$ requires higher-order Taylor approximations as more data become available; for Example 2, the mode is $k^*=4$ at both sample sizes, reflecting the smoother, monotone structure of the Michaelis--Menten curve.

\subsection{Summary}

The simulation confirms that DE1-$k$ estimation with automatic degree selection outperforms unconstrained local linear regression on both function estimation accuracy and ODE consistency, across both examples and both sample sizes considered, with the advantage particularly pronounced for MAE. These findings are consistent with the asymptotic bias reduction established in Theorem~\ref{thm:bias} and complement the application results of Section~\ref{sec:application}.

\section{Discussion}
\label{sec:discussion}

The theoretical, simulation, and application results in this paper are mutually consistent. Theorem~\ref{thm:bias} shows that the DE1-$k$ estimator reduces asymptotic bias at a rate governed by the Taylor degree $k$, without inflating the asymptotic variance (Theorem~\ref{thm:variance}); the simulation study confirms this in finite samples for two qualitatively different ODE structures; and the Albini application shows the same pattern in real, sparse experimental data, where the DE-constrained estimator's advantage over local linear regression is concentrated in the two smallest groups ($n=3$ and $n=4$) and disappears once $n\geq5$. Taken together, these results support the view that DE-constrained LPR is most useful precisely when data are too sparse for unconstrained smoothers to be reliable and a credible mechanistic model is available to fill the gap, which is a common situation in laboratory fire science, where experiments are costly and sample sizes per condition are small.

Section~\ref{sec:misspecification} shows that this benefit is conditional on the constraint being approximately correct: an LQL constraint applied to LQE-generated data introduces a bias term that does not vanish with bandwidth or sample size, which is the price of building physical structure into the estimator. This motivates two directions for future work. First, the two firebrand transport models discussed alongside the burning-rate model in Appendix C of \citet{albini1979spot}, a nonparametric wind-tunnel model based on $d(\rho_s D)/dt = -m'(t)\rho_a u$ and a nonparametric spotting-fire model based on $d(\rho_s D)/dt = -m'(t)\rho_a v_o(t)$ with $v_o(t)=\sqrt{\pi g(\rho_s D)/(2C_D\rho_a)}$, are natural extensions of the present framework to the downstream problem of predicting spotting distance itself, rather than burning rate alone. Second, combining the DE constraint with the variance-function bias-reduction technique of \citet{cheng2018bias}, and extending the varying-coefficient formulation $g(t;X)=a_1(t)+a_2(t)X$ to model species as a covariate directly rather than by stratification, would allow the eight species-diameter groups analyzed separately here to be pooled into a single model, at the cost of additional modelling complexity.

A limitation of the current framework is that the choice of degree $k$ and bandwidth $h$, while guided by the MAE/MADE criteria in simulation and by the pilot plug-in bandwidth in the application, remains partly heuristic; a fully data-driven joint selection rule for $(h,k)$ is left for future work. The two approaches to estimating $a(x)$ and $b(x)$ in Section~\ref{sec:estimate-ab} also rely on a pilot local-constant fit, whose own error propagates into the final DE-constrained estimate; characterizing this propagation formally would strengthen the asymptotic theory of Section~\ref{sec:theory} for the case of unknown $(a,b)$.

\section{Conclusions}
\label{sec:conclusions}

This paper extended differential equation-constrained local polynomial regression from the exponential growth model of \citet{ge2026differential} to the general first-order linear ODE $g'(x)=a(x)g(x)+b(x)$, which we call the local quasi-linear (LQL) model. We derived closed-form DE1-$k$ estimators for arbitrary Taylor degree $k$, established their asymptotic conditional bias and variance, and proposed two approaches for estimating $a(x)$ and $b(x)$ when they are unknown. A simulation study across two structurally different ODEs confirmed that DE1-$k$ estimation with automatic degree selection reduces both estimation error and ODE-consistency error relative to unconstrained local linear regression. Applied to the Albini (1979) firebrand burning-rate data, the LQL model, with $a(t)$ and $b(t)$ identified from a physically motivated forced-convection ODE, improved on local linear regression in the sparsest experimental groups, where physical structure is most valuable in compensating for scarce data. We also examined the robustness of the LQL estimator to misspecification against a local quasi-exponential alternative, showing that the benefit of the ODE constraint depends on its being approximately correct. Together, these results extend the DE-constrained regression paradigm to a broad class of physically motivated linear models and provide a practical estimation tool for fire science and other application areas where mechanistic knowledge is available but data are sparse.

\section*{Acknowledgements}

The simulation code used in Section~\ref{sec:simulation} was originally developed by W.\ J.\ Braun and subsequently adapted by C.\ Ge. This research has been supported in part by a grant from the Natural Sciences and Engineering Research Council of Canada (NSERC).

\bibliographystyle{apalike}
\bibliography{references}

\appendix
\section{Proofs of Theorems~\ref{thm:bias} and~\ref{thm:variance} for $k=1$}
\label{app:proofs}

\subsection{Proof of Theorem~\ref{thm:bias} ($k=1$)}

From~(\ref{equ:wlslinear1}), the conditional expectation of $\widehat{g}_1(x)$ is
\begin{align}
\mathrm{E}[\widehat{g}_1(x)\mid x_1,\ldots,x_n]
&= \frac{\sum_{i=1}^n (\mathrm{E}[y_i\mid x_1,\ldots,x_n]-(x_i-x)b(x))(1+(x_i-x)a(x))K_h(x_i-x) }{\sum_{i=1}^n (1+(x_i-x)a(x))^2K_h(x_i-x)} \nonumber\\
&= \frac{\sum_{i=1}^n (g(x_i)-(x_i-x)b(x))(1+(x_i-x)a(x))K_h(x_i-x) }{\sum_{i=1}^n (1+(x_i-x)a(x))^2K_h(x_i-x)} \nonumber\\
&\approx  \frac{\sum_{i=1}^n \bigl(g(x)+(x_i-x)g'(x)+\tfrac{1}{2}(x_i-x)^2g''(x)-(x_i-x)b(x)\bigr)(1+(x_i-x)a(x))K_h(x_i-x) }{\sum_{i=1}^n (1+(x_i-x)a(x))^2K_h(x_i-x)} \nonumber\\
&= g(x) + \frac{1}{2}\frac{\sum_{i=1}^n (x_i-x)^2\bigl((a'(x)+a^2(x))g(x)+a(x)b(x)+b'(x)\bigr)(1+(x_i-x)a(x))K_h(x_i-x) }{\sum_{i=1}^n (1+(x_i-x)a(x))^2K_h(x_i-x)},
\label{equ:EDE11}
\end{align}
where the second-order Taylor expansion of $g(x_i)$ uses $g'(x)=a(x)g(x)+b(x)$ and $g''(x)=(a'(x)+a^2(x))g(x)+a(x)b(x)+b'(x)$ from~(\ref{eqn:derivativeLinear}). Replacing the sums in~(\ref{equ:EDE11}) by integrals against the design density $f$, substituting $w=(z-x)/h$, and applying the mean value theorem for integrals,
\begin{align}
\mathrm{Bias}(\widehat{g}_1(x)\mid x_1,\ldots,x_n)
&= \mathrm{E}[\hat{g}_1(x)\mid x_1,\ldots,x_n]-g(x) \nonumber\\
&\approx  \frac{1}{2}\frac{\int_a^b (z-x)^2\bigl((a'(x)+a^2(x))g(x)+a(x)b(x)+b'(x)\bigr)(1+(z-x)a(x))f(z)K_h(z-x)\,dz }{\int_a^b (1+(z-x)a(x))^2f(z)K_h(z-x)\,dz} \nonumber\\
&=  \frac{1}{2}\frac{\int h^2w^2\bigl((a'(x)+a^2(x))g(x)+a(x)b(x)+b'(x)\bigr)(1+hwa(x))f(x+hw)K(w)\,dw }{\int (1+hwa(x))^2f(x+hw)K(w)\,dw} \nonumber\\
&\approx \frac{1}{2}\bigl((a'(x)+a^2(x))g(x)+a(x)b(x)+b'(x)\bigr)h^2\mu_2+o_p(h^2),
\end{align}
which is~(\ref{eqn:bias_k1_ex}) with $\mu_2=\int w^2K(w)\,dw<\infty$. The general-$k$ result of Theorem~\ref{thm:bias} follows by the same argument applied to the degree-$k$ Taylor approximant $g_k^*(x_i)$ built from the recursion~(\ref{eqn:derivativeLinear}), with the odd/even distinction arising from whether the leading term of the Taylor remainder is proportional to $\mu_{k+1}$ (which is zero for symmetric kernels when $k+1$ is odd) or to $\mu_{k+2}$.

\subsection{Proof of Theorem~\ref{thm:variance} ($k=1$)}

Under assumption~(IV), $\mathrm{Var}(y_i\mid x_1,\ldots,x_n)=\sigma^2(x_i)$. From~(\ref{equ:wlslinear1}),
\begin{align}
\mathrm{Var}(\widehat{g}_1(x)\mid x_1,\ldots,x_n)
&= \frac{\sum_{i=1}^n \mathrm{Var}(y_i\mid x_1,\ldots,x_n)(1+(x_i-x)a(x))^2K_h^2(x_i-x) }{\{\sum_{i=1}^n (1+(x_i-x)a(x))^2K_h(x_i-x)\}^2}\nonumber\\
&= \frac{\sum_{i=1}^n \sigma^2(x_i) (1+(x_i-x)a(x))^2K_h^2(x_i-x) }{\{\sum_{i=1}^n (1+(x_i-x)a(x))^2K_h(x_i-x)\}^2} \nonumber\\
&\approx \frac{\sum_{i=1}^n \bigl( \sigma^2(x)+(x_i-x)\tfrac{d}{dx}\sigma^2(x)\bigr) (1+(x_i-x)a(x))^2K_h^2(x_i-x) }{\{\sum_{i=1}^n (1+(x_i-x)a(x))^2K_h(x_i-x)\}^2},
\end{align}
using a first-order Taylor expansion of $\sigma^2(x_i)$ about $x$ (assumption~(IV)). Replacing sums by integrals and substituting $w=(z-x)/h$,
\begin{align}
\mathrm{Var}(\widehat{g}_1(x)\mid x_1,\ldots,x_n)
&\approx \frac{1}{n} \frac{\int_a^b \sigma^2(x) (1+(z-x)a(x))^2K_h^2(z-x)f(z)\,dz }{\{\int_a^b (1+(z-x)a(x))^2K_h(z-x)f(z)\,dz\}^2}\nonumber \\
&= \frac{1}{nh} \frac{\int \sigma^2(x) (1+hwa(x))^2K^2(w)f(x+hw)\,dw }{\{\int (1+hwa(x))^2K(w)f(x+hw)\,dw\}^2}\nonumber \\
&\approx \frac{\sigma^2(x)R(K)}{nhf(x)}+o_p\!\left(\frac{1}{nh}\right).
\end{align}
The general-$k$ result of Theorem~\ref{thm:variance} follows by the same argument with $(1+(x_i-x)a(x))$ replaced by the degree-$k$ analogue from~(\ref{eqn:derivativeLinear}); the leading-order variance is unchanged because the additional Taylor terms are $O(h)$ and enter only the $o_p(1/(nh))$ remainder.

\label{lastpage}
\end{document}